\documentclass[pr12pt]{iopart}
\usepackage{bbold}
\usepackage{graphicx, caption}
\usepackage{iopams}
\expandafter\let\csname equation*\endcsname\relax
\expandafter\let\csname endequation*\endcsname\relax
\usepackage{amssymb}
\usepackage{amsmath}
\usepackage{empheq}
\usepackage{amsthm}
\usepackage{float}
\usepackage{siunitx}

\usepackage{color}
\usepackage{lineno}

\begin{document}

\title[Rainbow Trapping and Rainbow Reflection]{
Rainbow reflection and trapping for 
energy harvesting}

\author{G.~J. Chaplain \textsuperscript{1}, Daniel Pajer \textsuperscript{2}, Jacopo M. De Ponti \textsuperscript{3,4} \& R.~V. Craster \textsuperscript{1,5}}

\ead{gregory.chaplain16@imperial.ac.uk}
\address{$^1$ Department of Mathematics, Imperial College London, 180 Queen's Gate, South Kensington, London SW7 2AZ}
\address{$^2$ Department of Physics, The Blackett Laboratory, Imperial College, London SW7 2BZ, United Kingdom}
\address{$^3$Department of Civil and Environmental Engineering, Politecnico di Milano,
	Piazza Leonardo da Vinci, 32, 20133 Milano, Italy}
\address{$^4$Department of Mechanical Engineering, Politecnico di Milano,
	Via Giuseppe La Masa, 1, 20156 Milano, Italy}
\address{$^5$Department of Mechanical Engineering, Imperial College London, London SW7 2AZ, UK}

\begin{abstract}
Important distinctions are made between two related wave control mechanisms that act to spatially separate frequency components; these so-called rainbow mechanisms either slow or reverse guided waves propagating along a graded line array. We demonstrate an important nuance distinguishing rainbow reflection from genuine rainbow trapping and show the implications of this distinction for energy harvesting designs. The difference between these related mechanisms is highlighted using a design methodology, applied to flexural waves on mass loaded thin Kirchhoff-Love elastic plates, and emphasised through simulations for energy harvesting in the setting of elasticity, by elastic metasurfaces of graded line arrays of resonant rods atop a beam. The delineation of these two effects, reflection and trapping, allows us to characterise the behaviour of forced line array systems and predict their capabilities for trapping, conversion and focusing of energy. 
\end{abstract}

\section{Introduction}
\label{sec:Intro}
Graded line arrays capable of supporting array guided waves have recently been theorised, designed, and manufactured with energy harvesting capabilities at the forefront of the proposed applications \cite{Alan2019,DePonti2019}. These simple devices are often based around gradually varying periodic arrays to take advantage of local band-gaps, or of waveguides of varying thicknesses, to control wave propagation; array guided waves slow down as they transverse the array with different frequency components localising at specific spatial positions, resulting in a so-called `trapped rainbow'. This effect originated in electromagnetism being predicted and observed using axially non-uniform, linearly tapered, planar waveguides with cores of negative index material \cite{Tsakmakidis2007}. Subsequently, this effect has been achieved in a host of wave regimes, without explicit analogue to a negative index, where a surface is structured by discrete, often subwavelength, unit cells comprising resonant elements. Such metamaterials and metasurfaces have achieved so-called trapping and enhancement effects for spoof surface plasmon polaritons \cite{Gan2008,Gan2011}, acoustics \cite{Romero-Garcia2013,Cebrecos2014}, water waves \cite{bennetts18a} and fluid loaded elastic plates \cite{skelton2018multi} with particular advances in elastic metawedge devices using arrays of resonant rods, for both thin elastic plates and deep elastic substrates \cite{Alan2019,DePonti2019,colombi16a,Colombi2017,Chaplain2019a}. 

The underlying physics, fundamental to the design of graded structures capable of spatial segregation of frequency components, for either trapping or mode conversion applications, relies on the ability to isolate the dispersion curves of the locally periodic structures which make up such an array. A graded array is formed by gently, i.e. adiabatically, varying a particular parameter, or set of parameters of neighbouring elements in subsequent unit cells. Physically, provided the grading is gentle enough, the global behaviour of the whole array is deduced from the knowledge obtained from the local dispersion curves of the constituent elements \cite{colombi16a}; the desired spatial selection by frequency properties, the rainbow behaviour, of the device is determined from the locally periodic structure at a given position. For example, in the case of conventional elastic metawedge devices, resonant rods are fixed atop a thin elastic plate or deep substrate such as a half-space; when local band gaps are reached at say, a certain rod height, the inability of the wave to propagate beyond this position can be leveraged to produce a variety of effects, depending on the operating frequency \cite{Chaplain2019b}.

\begin{figure}
    \centering
    \includegraphics[width = 0.95\textwidth]{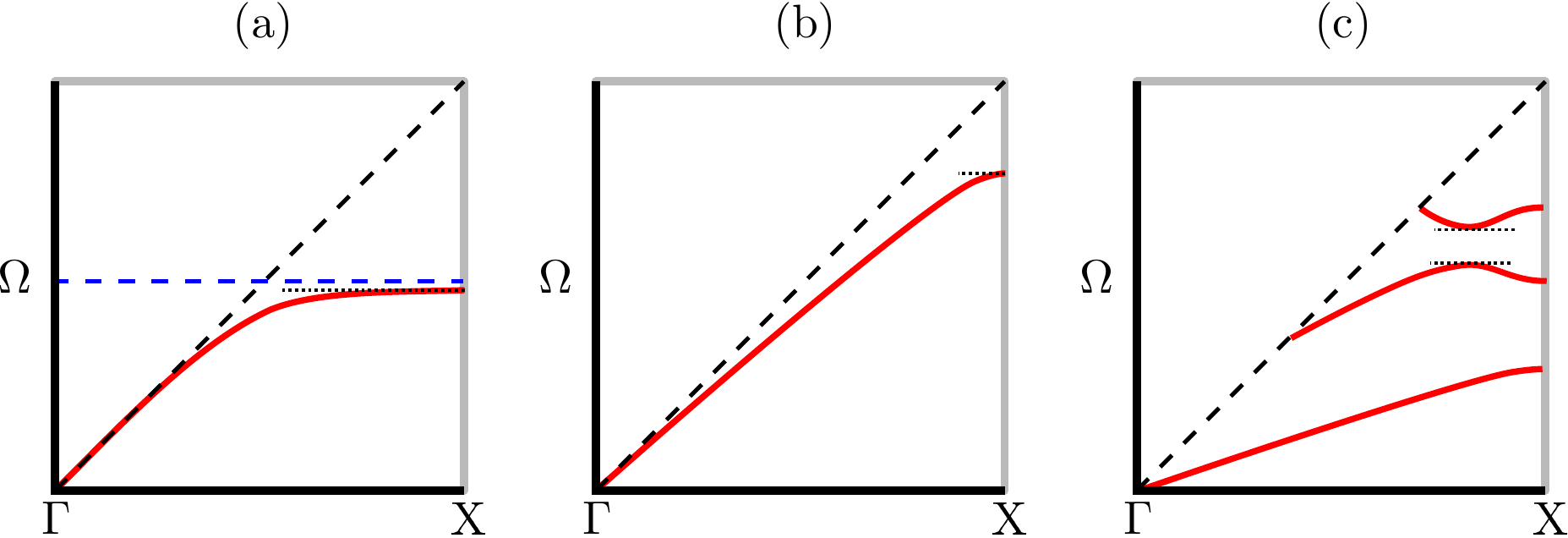}
    \caption{Dispersion curves for differing periodic systems, with zero group velocity modes shown with dotted black lines. (a) Typical resonant system with zero group velocity mode induced by resonance (dashed blue line). (b) A (potentially non-resonant) system where zero group velocity modes are induced by the Bragg condition only, i.e. due to the periodicity. (c) Symmetry broken system, where accidental degeneracies are lifted resulting in zero group velocity modes inside the Brillouin Zone.}
    \label{fig:CompCurves}
\end{figure}

In adiabatically graded arrays, the interpretation of the locally periodic dispersion curves has led to a conflation between rainbow trapping phenomena and other related slow wave effects. Ultimately, `trapping' of array guided waves has been accounted for via the drastic reductions in the group velocity of the waves which transit the array, eventually reaching a region with zero group velocity. With the exception of acoustic analogues which discuss `soft reflections' \cite{Romero-Garcia2013,Cebrecos2014}, many graded systems ignore any reflections from these regions. Our purpose here is to draw out the nuanced difference between simply slowing down the waves and genuinely trapping them. 

These rainbow devices emerged from concepts of slow light materials, that exploit resonances in dispersive materials \cite{VestergaardHau1999}. Similar effects were realised using (line-defect) photonic crystal waveguides \cite{Notomi2001,Vlasov2005,Baba2008}, which exactly exploit unique dispersive properties resulting in zero group velocity modes. The introduction of negative index materials and tapering of the waveguide in \cite{Tsakmakidis2007} achieved a similar broadband result and first introduced the term `trapped rainbow', focusing on regions of zero effective thickness, as opposed to zero group velocity modes. Subsequent translations of this effect to other graded wave systems, which do not use negative index materials, have attributed \textit{all} zero group velocity modes with the ability to perform rainbow trapping. 
We investigate zero group velocity modes in detail and show that they are not all the same, and that some reflect energy all be it slowly and others genuinely trap energy. We  analyse an elastic system, in the frequency domain, taking into account reflections by local standing wave modes where these zero group velocity modes arise. We show that in the long time limit the origin of these zero group velocity modes becomes important in delineating trapping and reflecting phenomena.

Zero group velocity modes used to infer the rainbow trapping phenomenon in discrete graded systems, can arise through a number of avenues. Shown in Fig.~\ref{fig:CompCurves} are the typical dispersion curves for three model systems were zero group velocity modes exist. Figure~\ref{fig:CompCurves}(a) shows a flat dispersion branch, typical of slow sound or slow light in periodic systems composed of resonant elements, where over a large region in wavenumber space the group velocity is very low. Such a system is analogous to original slow light devices where, over a narrow frequency range, the group velocity can be greatly reduced compared to a free space wave. Slow sound acoustic analogues have utilised this effect in systems of Helmholtz resonators, whereby including losses into the system `trapping' occurs via absorption \cite{Jimenez2017a,Jimenez2017b}. Figure~\ref{fig:CompCurves}(b) shows the dispersion for a (potentially non-resonant) periodic array, where a zero group velocity mode is achieved purely due to the Bragg condition being met at the band edge; in a perfectly periodic medium standing waves form at this frequency due to \textit{reflections}. These reflections prove key to delineating between true rainbow trapping and what we will term `rainbow reflection' phenomena. In Fig.~\ref{fig:CompCurves}(c) we show the dispersion curves for a symmetry broken array, where an accidental degeneracy is lifted, resulting in zero group velocity modes \textit{within} the first Brillouin Zone. We will focus on geometries capable of supporting such modes, which offer larger trapping potential due to the lack of coupling with reflected modes. 

Whilst the positions of the zero group velocity modes in reciprocal space may seem a technicality, we show these nuances and their effect on the resulting wave phenomena have strong influences on energy harvesting applications for such arrays. We first focus on a model of graded line arrays of point masses on thin Kirchhoff--Love (KL) elastic plates, to delineate between true rainbow trapping, and the related rainbow reflection which have, until now, been used interchangeably. This KL model is attractive due to the fast, accurate spectral methods available to identify the local dispersion curves of the singly periodic structures \cite{Chaplain2019a} which support Rayleigh-Bloch modes. The KL model is popular in elastic wave physics in part for technical reasons, the Green's function is non-singular \cite{evans2007penetration} allowing for fast and accurate numerical simulation \cite{xiao2012flexural},  
also because in its range of applicability it captures experimental behaviour accurately \cite{lefebvre17a},
 and furthermore as it allows for the insightful exploration of topological wave physics and edge states \cite{Pal2017,torrent13a}.  
 The Rayleigh-Bloch modes are array guided waves that propagate along the direction of the array, and decay exponentially perpendicular to the array  \cite{evans2007penetration}. An example of a line array capable of exhibiting both rainbow trapping and reflection (amongst other effects) is outlined in Section~\ref{sec:LineArrays}, along with the general design paradigm to achieve true rainbow trapping. To further illustrate the importance of the difference between these two effects, we propose a new piezo-augmented array for harvesting electric energy from elastic substrates, and compare its functionalities for ungraded arrays and graded arrays capable of employing rainbow trapping and rainbow reflection.


\section{Graded Line Arrays}
\label{sec:LineArrays}

Recent studies of ungraded line arrays created using clusters of resonators placed on thin elastic plates \cite{packo19a} have illustrated the power of using array systems in terms of manipulating the transmission of flexural elastic waves. We use similar arrays, although we can illustrate the concepts required here using simpler mass-loaded line arrays ultimately with the added ingredient of grading their properties along the array. 

\begin{figure}
	\centering
	\includegraphics[width=0.99\textwidth]{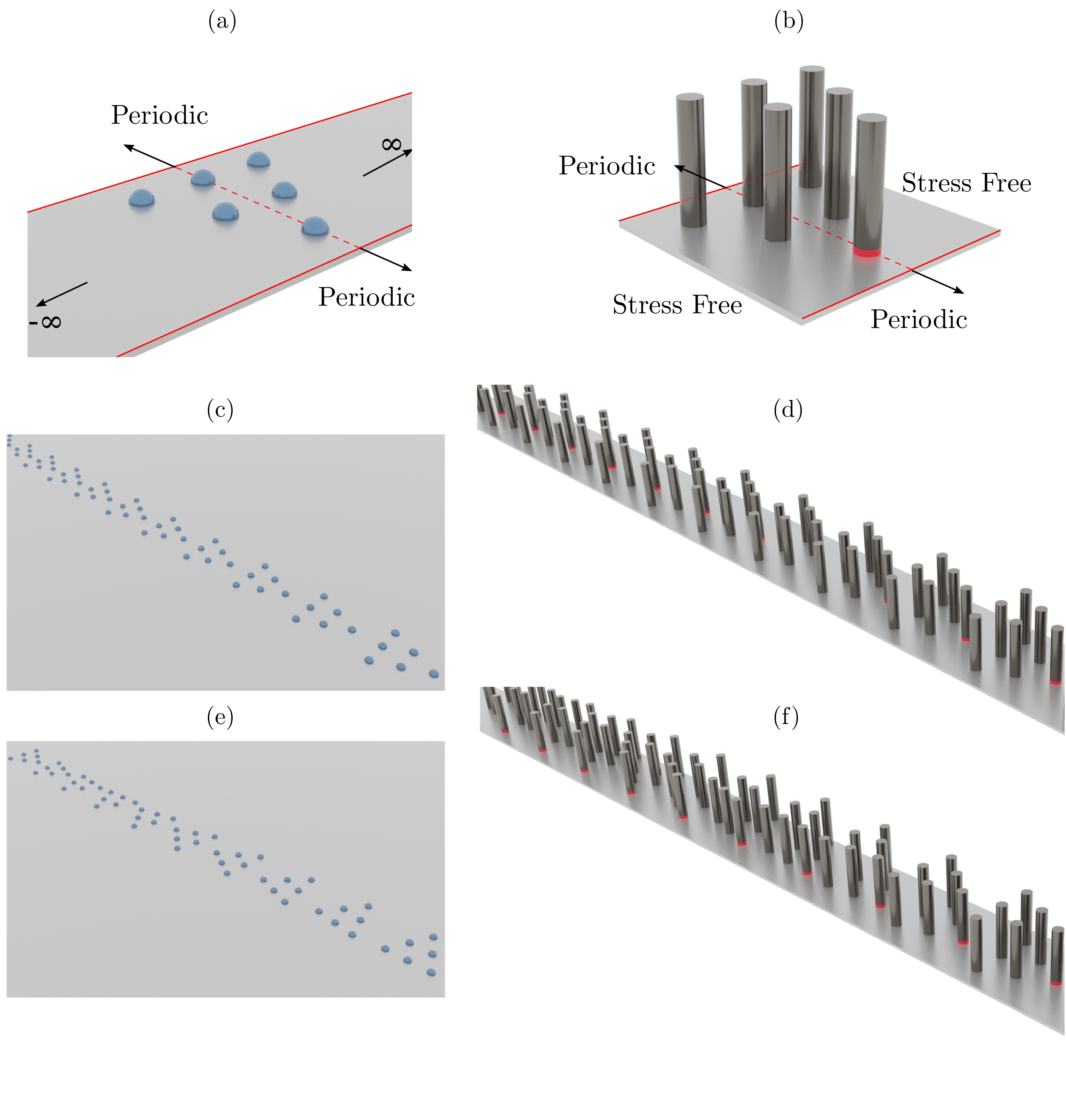}
	\caption{(a) Unit strip (solid red lines) that periodically repeats and which has a cluster point masses, arranged to form a triangle, represented by blue half spheres and with the array axis shown as dashed red line. (b) Unit cell with a cluster of resonant rods placed atop an elastic beam with boundary conditions as shown. A piezopatch (red material) attached to a single rod is shown for harvesting applications. (c,d) show ungraded arrays of masses on a KL plate and rods atop a beam respectively, with corresponding examples of graded arrays, where the grading is introduced by rotation of the triangular geometries, shown in (e,f).}
	\label{fig:schem}
\end{figure}

To demonstrate the key difference between rainbow trapping and rainbow reflection, we first analyse a line array of point masses clustered into triangular arrangements placed atop a thin KL elastic plate and ignore grading in order to understand the perfectly periodic system. The line array is one-dimensional in the sense that it is singly periodic, relative to the array axis (Fig.~\ref{fig:schem}). To generate the dispersion curves of this system, and of the subsequent graded systems, we partition the array into infinite unit strips, so that the governing equation for the out-of-plane flexural wave displacement, $w(\boldsymbol{x})$, is \cite{Chaplain2019a}

\begin{equation}
\nabla^{4}w(\mathbf{x}) - \Omega^{2}w(\mathbf{x}) = \Omega^2\sum\limits_{N,j = 1}^{J} M_{N}^{(j)}w(\mathbf{x})\delta(\mathbf{x}-\mathbf{x}_{N}^{(j)}),
\label{eq:KL}
\end{equation}
where we utilise a nondimensionalised frequency such that \(\Omega^2 = \rho h\omega^2/D\), where \(\rho\) is the mass density of the plate and \(h\) is the plate thickness, with \(\omega\) being the dimensional angular frequency. \(D\) is the flexural rigidity, which encodes the Young's modulus, \(E\), and Poisson's ratio, \(\nu\), of the plate through \(D = Eh^3/12(1-\nu^2)\). The unit strips are labelled $N$, with there being $J$ masses within each strip at a position $\mathbf{x}_{N}^{(j)}$. An example unit strip schematic with 6 masses arranged in a triangular geometry is shown in Fig.~\ref{fig:schem}(a), with a section of a perfectly periodic line array shown in Fig.~\ref{fig:schem}(c).

\begin{figure}
	\centering
	\includegraphics[width=0.99\textwidth]{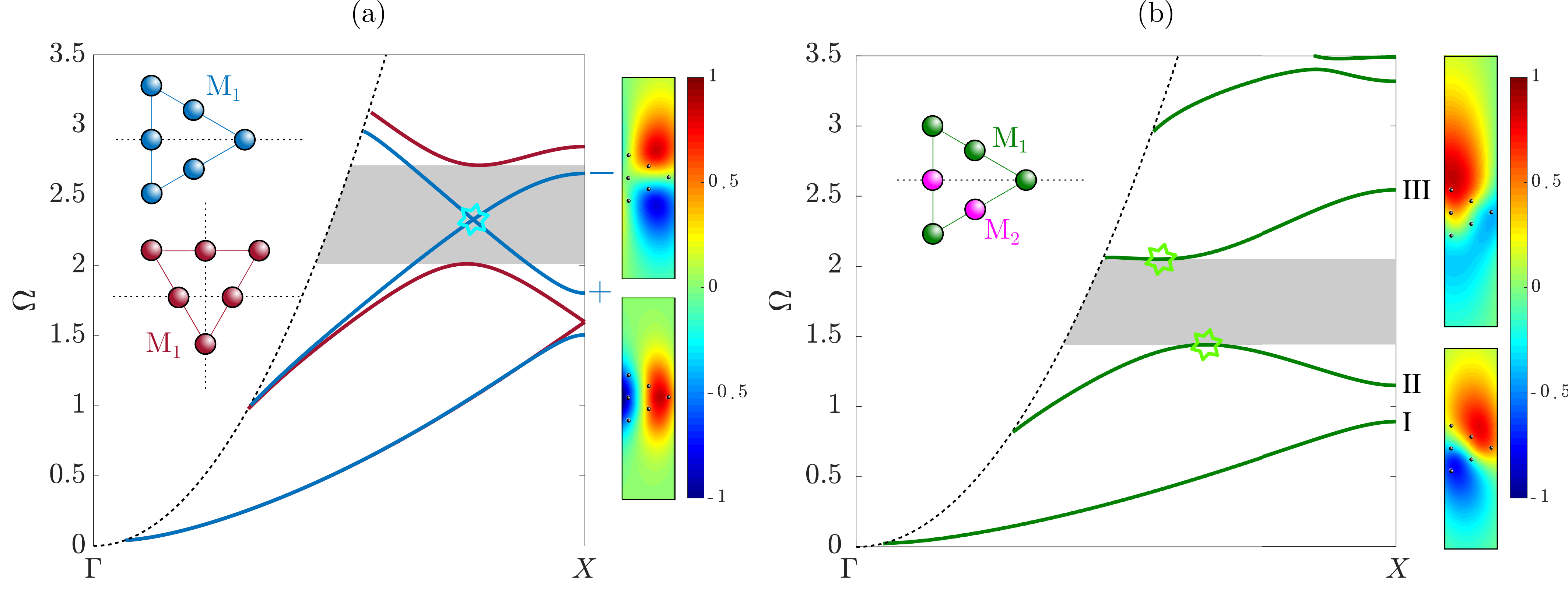}
	\caption{(a) Dispersion curves for array with symmetry of axis (blue) showing accidental degeneracy, highlighted by blue star. Red curves show band gap opening due to reflectional symmetry with respect to array axis being broken. Insets show antisymmetric(odd) and symmetric(even) solutions at the degeneracy for the blue bands marked $(-,+)$ respectively, with $a=1$, $M=10$. (b) Symmetry broken curves by mass loading such that $M_{1} = 10$, $M_{2} = 5M_{1}$. Insets show wavefields at minima of band III and maxima of band II respectively (marked by green stars). The mixing of odd and even states can be seen through the varying amplitudes and decay lengths either side of the array axis.}
	\label{fig:Disp}
\end{figure}

The spectral method employed, as in \cite{Chaplain2019a}, rapidly and accurately characterises the dispersion curves of this structure. For a particular value of mass loading, $M = 10$, such curves are shown in blue in Fig.~\ref{fig:Disp}(a). In order to differentiate between the two effects, the symmetry of the inclusions within the unit cell is of paramount importance; rainbow reflection effects can be achieved with \textit{any} inclusion geometry, since ultimately they leverage only the Bragg condition by virtue of the periodicity. For true rainbow trapping effects, trapping must be located at wavevectors \textit{within} the first Brillouin Zone (BZ), and hence rely on the decoupling of orthogonal eigensolutions, or symmetry breaking of the array geometry \cite{makwana2018geometrically,Makwana2019tuneable} so to lift accidental degeneracies within the first BZ. 

\begin{figure}
	\centering
	\includegraphics[width=0.95\textwidth]{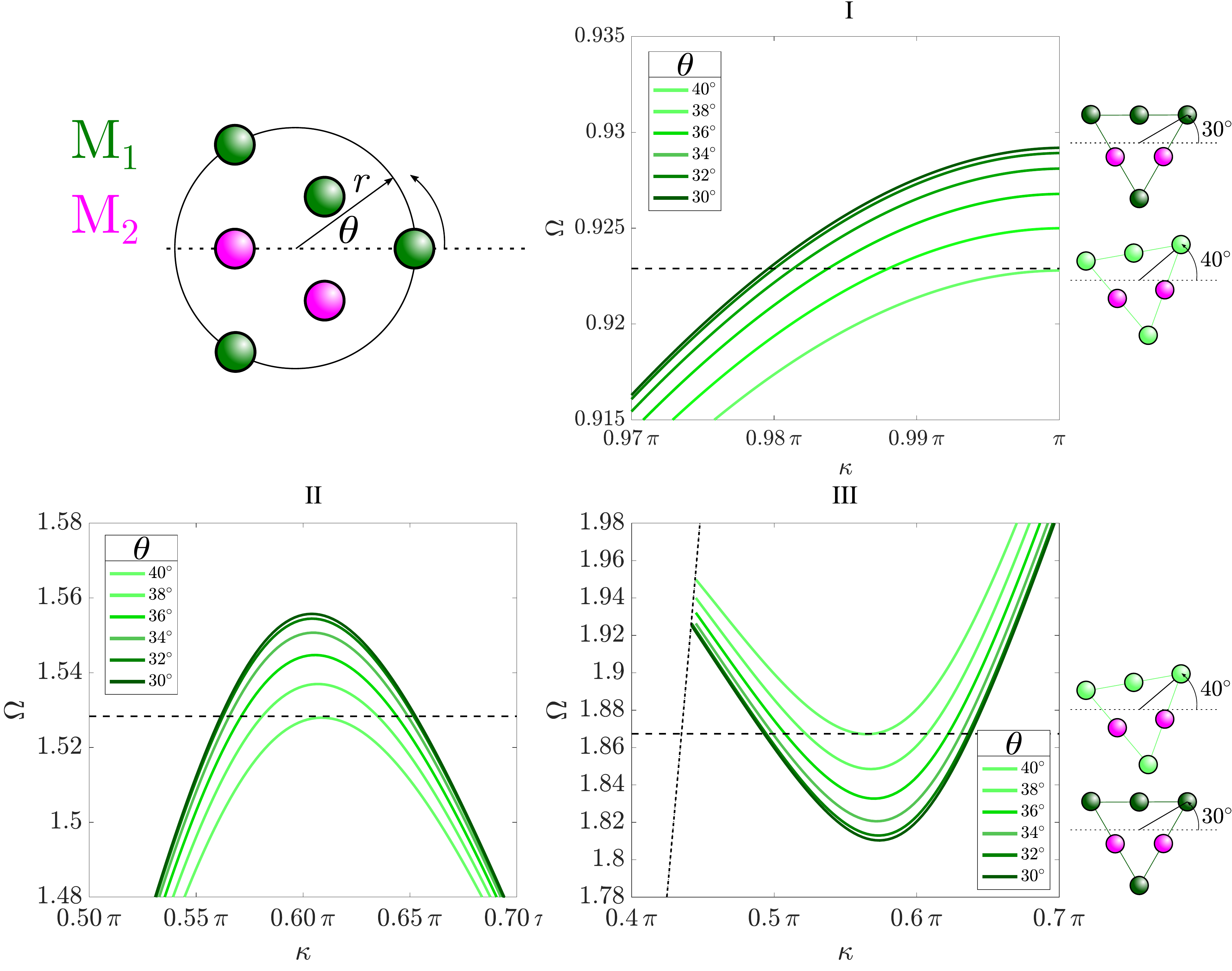}
	\caption{Effect of rotating asymmetric mass loaded structure, showing zoom on bands I,II and III. Horizontal dashed lines show the frequencies stopped at inclusion angles of $\theta = 40^{\circ}$. Whether trapping or reflection subsequently occurs depends on the wavevector at which $v_{g} = 0$ is achieved; $\Omega = 0.923$ lies on band I and exhibits rainbow reflection at this angle, whilst $\Omega = 1.528, 1.867$ lie on bands II and III which exhibit rainbow trapping.}
	\label{fig:GradingDisp}
\end{figure}

Central to designing rainbow trapping and reflecting line arrays, is to adiabatically grade the array with respect to some set of parameters, thereby altering the local dispersion curves of the structure. The choice of grading parameter can be any combination of the inclusion geometry, mass values, unit cell or loading element (for KL plates these can be pins, masses or resonators \cite{evans2007penetration}). By altering such parameters, the dispersion curves of subsequent unit strips are pushed up or down in frequency, or indeed completely change their shape. For a given frequency, there will then be regions were a wave is either supported or prohibited from propagating; local band gaps are then reached at different spatial positions for different frequency components. Conventional elastic, acoustic and water wave systems have focused on how the \textit{lowest} dispersion branch is changed via the grading parameter, maintaining the symmetry properties of the array. Most of these systems can then only exhibit reflective effects; higher order modes need to be present in the presence of a symmetry broken system for true trapping for discrete arrays. This insight motivated the choice of the triangular mass loading geometry since it supports both purely odd and even modes with respect to the array axis in the symmetric configuration, and modes which are neither odd nor even when symmetry is broken. The analysis presented is completely general and holds so long as accidental degeneracies within the BZ can be exploited. When analysing the changes of the lowest dispersion branch, it is possible to engineer a local band gap where, at some unit cell for a given frequency, the resultant graded wavevector corresponds to the band edge, that is the boundary of the first Brillouin Zone (Fig.~\ref{fig:GradingDisp}). At this position in a conventional perfectly periodic system standing waves form through subsequent Bragg reflections due to the periodicity. In graded systems, this has been labelled as rainbow trapping, since there is no forward propagating mode to couple into beyond this position, due to the encountered band gap. However, particularly in systems where no resonance effects are encountered \cite{Chaplain2019a}, this mode is quickly reflected and couples to a counter propagating wave which travels along the array in the opposite direction. Indeed the reflected wave after this `trapping' can be used for applications other than energy harvesting, such as flat lenses by passive self phased effects \cite{Chaplain2019b}. The misnomer of rainbow trapping has been attributed to this effect in almost all graded systems where locally periodic dispersion curves are analysed, as the wave is seen to be prohibited from propagating further along the array. Due to the reduced speed of the wave, it can appear to stay localised for a considerable length of time \cite{Gan2008}, however unlike a truly trapped wave, this wave will ultimately reflect due to the position where the group velocity vanished (at the BZB). This interpretation is corrected here, where we identify the apparent trapping of such a guided mode to be reflected via frequency domain analysis. The position of the reflection is frequency dependent through the grading parameter, and as such we term this corrected effect `rainbow reflection'. 

To distinguish between this reflection phenomenon and desired true rainbow trapping, we now utilise symmetry broken arrays. Degenerate eigensolutions of the dispersion relation, or accidental degeneracies (band crossings) within the Brillouin Zone (Fig.~\ref{fig:Disp}(a)) correspond to orthogonal modes and can exist if there is reflectional symmetry of the inclusion geometry about the array axis \cite{makwana2018geometrically}. These more complex geometries are not normally analysed in graded systems, but the extension to such geometries is trivial. Upon the breaking of the reflectional symmetry of the array, the degeneracy is lifted giving rise to solutions which are neither symmetric (even) or antisymmetric (odd) with respect to the array axis. For the array of triangular point masses, this symmetry breaking is achieved by simple rotations of the array or by altering mass values \cite{Makwana2019tuneable} in Fig.~\ref{fig:Disp}(a) and in Fig.~\ref{fig:Disp}(b), such that the reflectional array symmetry is broken. We will utilise these symmetry broken arrays to exploit the change in wavevector at which zero group velocity is achieved. A close up of an asymmetric arrangement of masses is shown in Fig.~\ref{fig:GradingDisp}. Here the grading parameter along the array will be the rotation of the array by an angle $\Delta\theta = 2\pi/N$, where $N$ is the number of unit cells within the graded region. In this way we can control the rate of grading and how quickly the dispersion curves change; Figure~\ref{fig:GradingDisp} shows the effect of this rotation on the locally periodic dispersion curves labelled I,II,III in Fig.~\ref{fig:Disp}(b). Rainbow reflection devices focus on manipulations similar to that of curve I; the wavevector with zero group velocity occurs at the band edge $\kappa = \pi/a$, where $a$ is the periodicity of the array. To prevent potentially undesirable reflection at the trapping position, we need to negate coupling into a backwards propagating mode. This is achieved by grading the symmetry broken arrays to create a zero group velocity mode that is encountered within the centre of the first BZ, as shown by the curves II and III in Fig.~\ref{fig:GradingDisp} (for a given frequency). If the grading is chosen such that the local band gap is encountered for a zero group velocity wavevector lying within the first BZ, then due to the lack of coupling to the reversed wavevector, true rainbow trapping can occur, resulting in vast field enhancement.

To characterise the differences between these two effects in this non-resonant system, we calculate the time averaged flux on a KL thin elastic plate through \cite{norris1995scattering}
\begin{equation}
    \langle \boldsymbol{F} \rangle = \frac{\Omega^2}{2}\mathfrak{Im}\left(w(\boldsymbol{x})\nabla^{3}w^{*}(\boldsymbol{x}) -\nabla^2w^{*}(\boldsymbol{x})\nabla w(\boldsymbol{x})\right),
\end{equation}
for an array in which neighbouring cells have geometries rotated by $\Delta\theta$. To determine the power radiated across each unit cell we integrate this quantity along the boundaries of each unit strip $x = ma$ with $m \in \mathbb{Z}$ such that
\begin{equation}
    I = \int\limits_{-\infty}^{\infty} \left(\langle \boldsymbol{F}\rangle \cdotp \hat{\mathbf{x}}\right) dy.
\end{equation}
These quantities are easily obtained from the frequency domain displacement fields calculated using a Green's function approach; this is one of the advantages of working with KL elastic plates since the Green's function is nonsingular and remains bounded \cite{evans2007penetration}.

\begin{figure}
	\centering
	\includegraphics[width=0.99\textwidth]{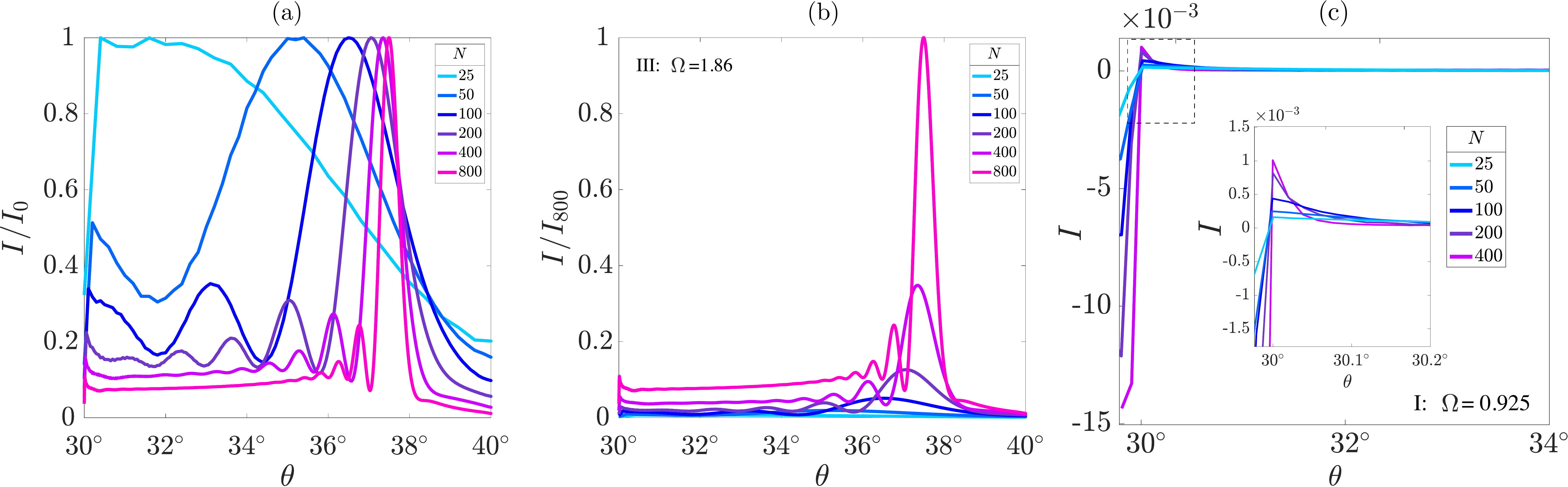}
	\caption{(a) Normalised power for asymmetric mass loaded array of Fig.~\ref{fig:GradingDisp}, graded with respect to $\Delta\theta = 2\pi/N$, normalised with respect to maximum power along each array, $I/I_{0}$, for $\Omega = 1.867$, corresponding to trapping along band III of Fig.~\ref{fig:GradingDisp}. This demonstrates the stronger confinement with decreasing grading parameter for rainbow trapping. (b) Similar to (a), but normalised to the finest array grading ($N = 800$), displaying the increase in localised amplitude with respect to grading parameter. (c) Similar plot to (a) but for $\Omega = 0.923$ for varying grading profile $\Delta\theta$. Here rainbow reflection can clearly be seen as almost zero power remains localised to the array; energy leaves the array, as can be seen by the negative power at the source position, corresponding to the rainbow reflection effect.}
	\label{fig:Flux}
\end{figure}

Shown in Fig.~\ref{fig:Flux} is the integrated flux, $I$, for bands I and III, for which rainbow reflection and trapping are clearly distinguished. In these simulations the array starts with $\theta = 30^{\circ}$ and is graded with varying values of $N$. Fig.~\ref{fig:Flux}(a,b) show that for trapping at a wavevector within the BZ, i.e. on band III, there is a large field enhancement near the trapping region, since the mode cannot couple into a reverse propagating mode. The effect of the rate of grading is pronounced; as $N$ is increased, the grading becomes more gradual, and as such the position of the grading is more localised, as shown in Fig.~\ref{fig:Flux}(a) which shows the normalised power with respect to each array. Further to this localisation, the amplitude of the trapped wave is also increased with decreasing $\Delta\theta$, as shown in Fig.~\ref{fig:Flux}(b) which shows the power normalised with respect to the largest $N$, and hence slowest grading. This confirms that true rainbow trapping is possible by trapping within the BZ, ensuring that there is minimal coupling to a reflected mode. In stark contrast to this, Fig.~\ref{fig:Flux}(c) shows the same calculations for the first band I as in Fig.~\ref{fig:GradingDisp}. For frequencies supported by this lowest branch, the grading parameter is almost irrelevant. Independently of where along the array the local band gap occurs, in the long time limit (as obtained by calculation of the fields in the frequency domain), all the energy is reflected; there is no field enhancement at the position of what has previously been referred to as `trapping'. The total power in this case is seen to be negative, i.e. in the opposite direction of the array origin (starting at $30^{\circ}$) indicating that all the energy leaks off the array after reflection for this non-resonant array. 

Having exemplified the differences between trapping and reflection, by the analysis of where the zero group velocity wave vector lies within the first BZ, we turn to a practical application of both these effects in the setting of energy harvesting in elastic media.

\section{Piezoelectric Applications}

Harvesting or scavenging vibrational energy is particularly attractive for the self powering of small electronic components, such as for the sensors used for structural health monitoring or medical implants \cite{Ahmed_2017}. In order to increase the efficiency of these devices, it is important to locally increase the vibrational energy that is already present in their immediate environment. This can be achieved by focusing, or trapping, waves from a larger region outside the device into a confined region in the near vicinity of the sensor; once the  wave is localized, by using electromagnetic, electrostatic or piezoelectric \cite{WILLIAMS19968} effects, efficient conversion from elastic to electric energy can be achieved.
When compared with other conversion methods, piezolectricity has the advantage of large power densities and ease of application \cite{Erturk2011,Erturk2013}, making it one of the most applicable energy harvesting methods. Several approaches have been employed in order to enhance harvesting efficiency, with properly designed structured materials or phononic crystals \cite{Carrara_2013, Gonnella2009}, lenses \cite{ErturkLens} and resonant metamaterials \cite{Sugino2018} also based on graded array designs \cite{Celli2016, DePonti2019}.

\begin{figure}
\hspace*{-0.2in}
    \centering
    \includegraphics[width = 1.08\textwidth]{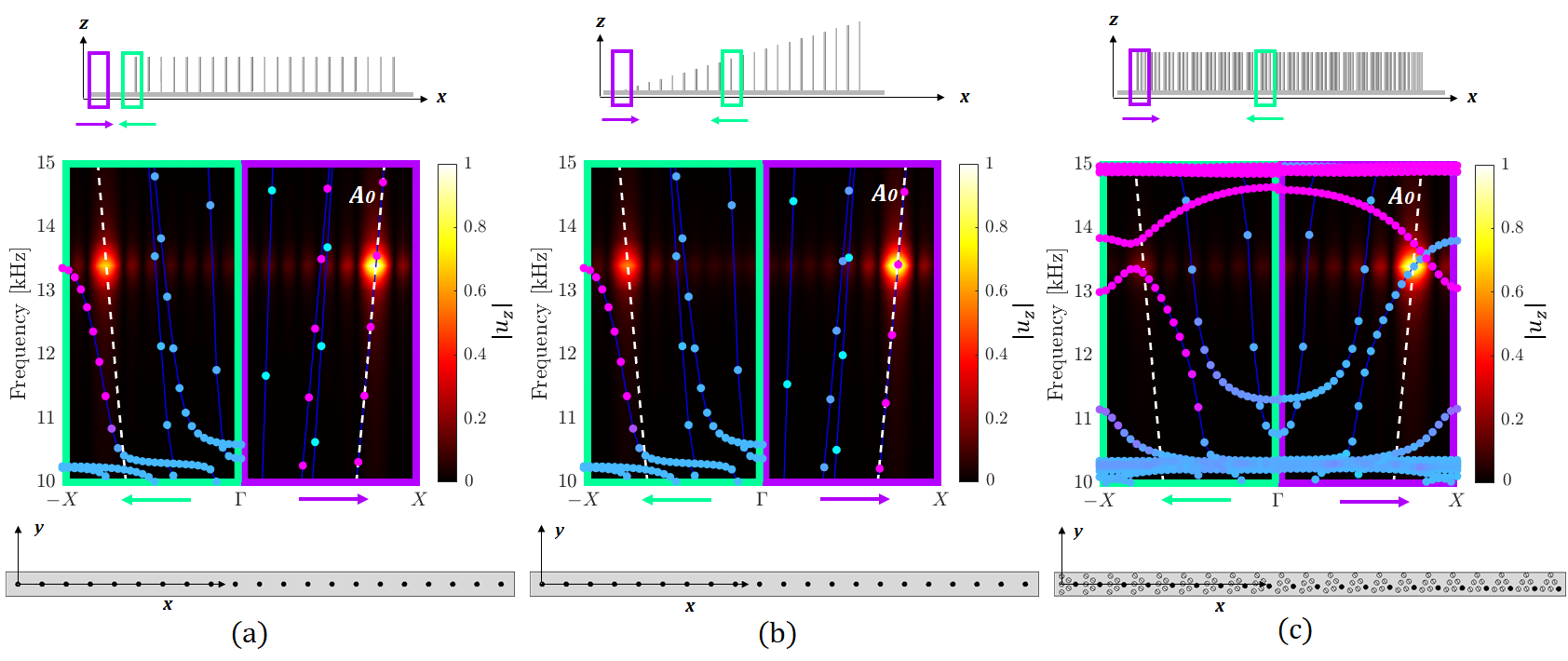}
    \caption{Dispersion spectra for array of constant height (a) a metawedge (b) and rotated cell array (c), corresponding to total scattering, Rainbow reflection and Rainbow trapping at time $10\si{\milli\second}$. Right panel shows spectra for the input wave, at position highlighted by purple rectangle in array schematics above, whilst the left panel shows the reflected wave spectra due to the bandgap opening, at cell positions matching green boxes. For (a,b) the bandgap opens via Bragg scattering, whilst in (c) due to the lifting of the accidental degeneracy. Overlaid on the spectra are the dispersion curves for highlighted cells; scatter points colours represent the wave polarization (purple corresponding to  vertical motion, i.e. axial elongation). The arrays are excited through an $A_0$ Lamb wave at $13.5\si{\kilo\hertz}$. Below each plot is an aerial view of the array.}
    \label{fig:10msPropagation}
\end{figure}

Motivated by our analysis, in section 2, on point mass loaded arrays, we adopt similar line arrays to compare the advantages of energy harvesting via true rainbow trapping and conventional rainbow reflective devices \cite{DePonti2019}. We compare two graded line arrays composed of clusters of aluminium rods ($E = 70\si{\giga\pascal}$, $\nu = 0.33$ and $\rho=2710\si{\kilo\gram\meter^{-3}}$) atop a beam, each with different gradings. The first exhibits conventional rainbow reflection; inspired by metawedge \cite{colombi16a, Chaplain2019d} structures we design a one dimensional array of rods, with a single rod per unit cell, each increasing in height in subsequent cells (Fig.~\ref{fig:10msPropagation}(b)). No symmetry induced accidental degeneracies arise in this setting, and as such to reach a zero group velocity mode, we must utilise the grading at the band edge by virtue of the Bragg condition. The second grading, incorporates the lifting of accidental degeneracies through breaking inversion symmetry through smoothly rotating a triangular array of rods within a unit cell from $0^{\circ}$ (i.e. symmetric about array axis) to $30^{\circ}$, similarly to the mass loaded plate analogy (Figs.~\ref{fig:schem}(b),~\ref{fig:10msPropagation}(c)). Both arrays are composed of $21$ cells with $30\si{\milli\meter}$ size and $10\si{\milli\meter}$ thickness. To quantify the reflection and trapping in the graded arrays, we analyse the dispersion spectra and compare both arrays with an ungraded periodic array of rods with equal number of unit cells. The ungraded array is composed of resonators of $80\si{\milli\meter}$ height and $3\si{\milli\meter}$ of diameter, similarly to  the array with rotated cells (see Fig. \ref{fig:10msPropagation}(a) and (c)). The metawedge (Fig. \ref{fig:10msPropagation} (b)) is obtained by a linearly grading defined by a $16^{\circ}$ slope angle, ranging in height from $2\si{\milli\meter}$ to $175\si{\milli\meter}$. Comparison is performed through a time domain simulation in Abaqus, exciting the line array for $30\si{\milli\second}$ with an antisymmetric (A0) Lamb wave with a central frequency of $13.5\si{\kilo\hertz}$, corresponding to the bandgap opening for rods of height $80\si{\milli\meter}$ and diameter $3\si{\milli\meter}$. Absorbing boundaries are imposed at the beam edges using the ALID method \cite{RAJAGOPAL201230}. Dispersion curves are computed in Abaqus with a user defined code capable of imposing Bloch-Floquet boundary conditions. The input and reflected waves are obtained by applying a spatiotemporal Fourier transform on the wavefield before the array. The reflected wave, as a percentage of the incident radiation, at $10\si{\milli\second}$ (Fig. \ref{fig:10msPropagation}) is $71\%$ for the ungraded array (a), $57\%$ for the rainbow reflective device (b)  and $21\%$ for the symmetry broken rainbow trapping configuration (c). There is then a clear difference in mechanism for the slowing down and reflecting/trapping for the respective arrays; the energy is stored for longer in the trapping device. 

\begin{figure}
\hspace*{-0.2in}
    \centering
    \includegraphics[width = 1.08\textwidth]{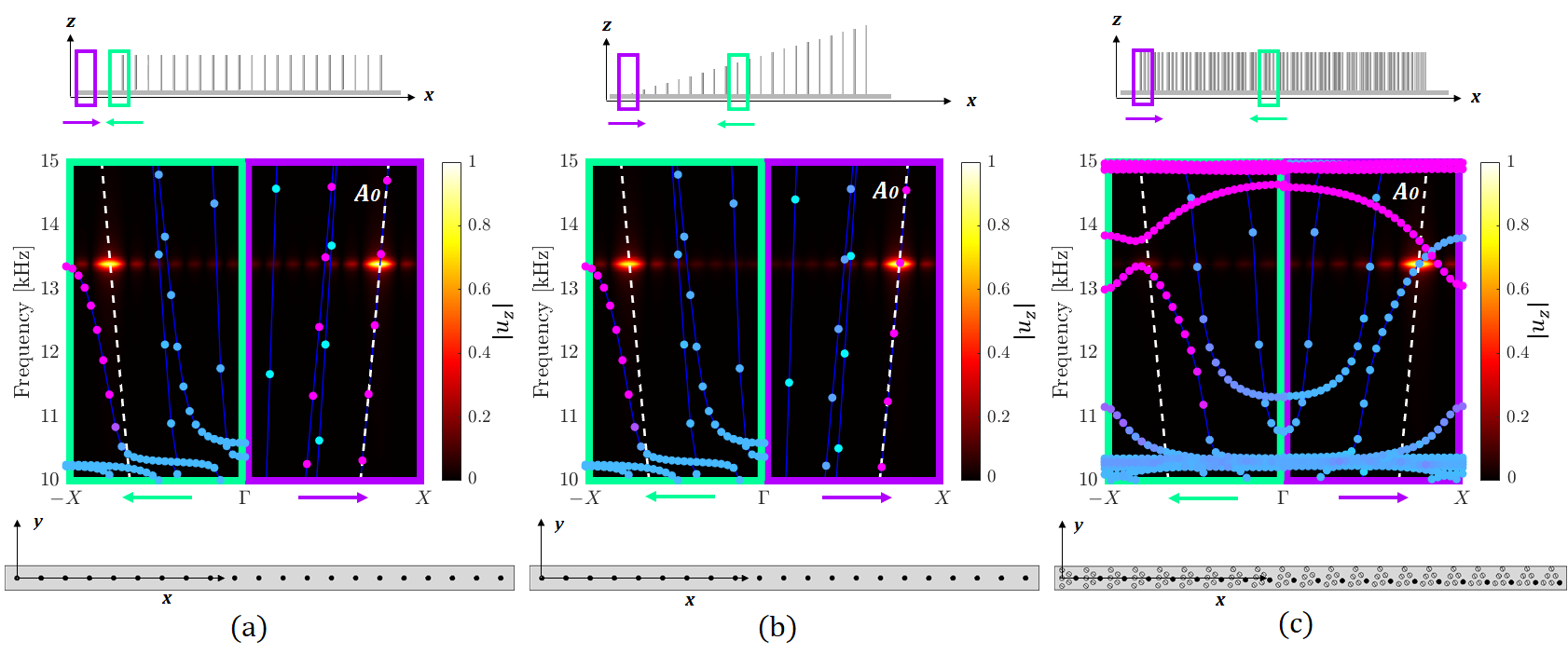}
    \caption{Similar to Fig.~\ref{fig:10msPropagation}, but at time of $30\si{\milli\second}$ (total input duration).}
    \label{fig:30msPropagation}
\end{figure}

This difference is emphasised when increasing the observation temporal window from $10$ to $30\si{\milli\second}$ (total input duration), we see the portion of reflected wave increases, reaching (Fig. \ref{fig:30msPropagation}), $91\%$ (a), $83\%$ (b) and $21\%$ (c) of the input signal for the scattering, reflection and trapping cases respectively. Considering reference to the case with equal rods (no grading), the metawedge reduces the reflections to almost $20\%$ at $10\si{\milli\second}$ and  $9\%$ at $30\si{\milli\second}$, while the rotating cell grading of $70\%$ at $10\si{\milli\second}$ and $77\%$ at $30\si{\milli\second}$. 

\begin{figure}
\hspace*{-0.2in}
    \centering
    \includegraphics[width = 1.08\textwidth]{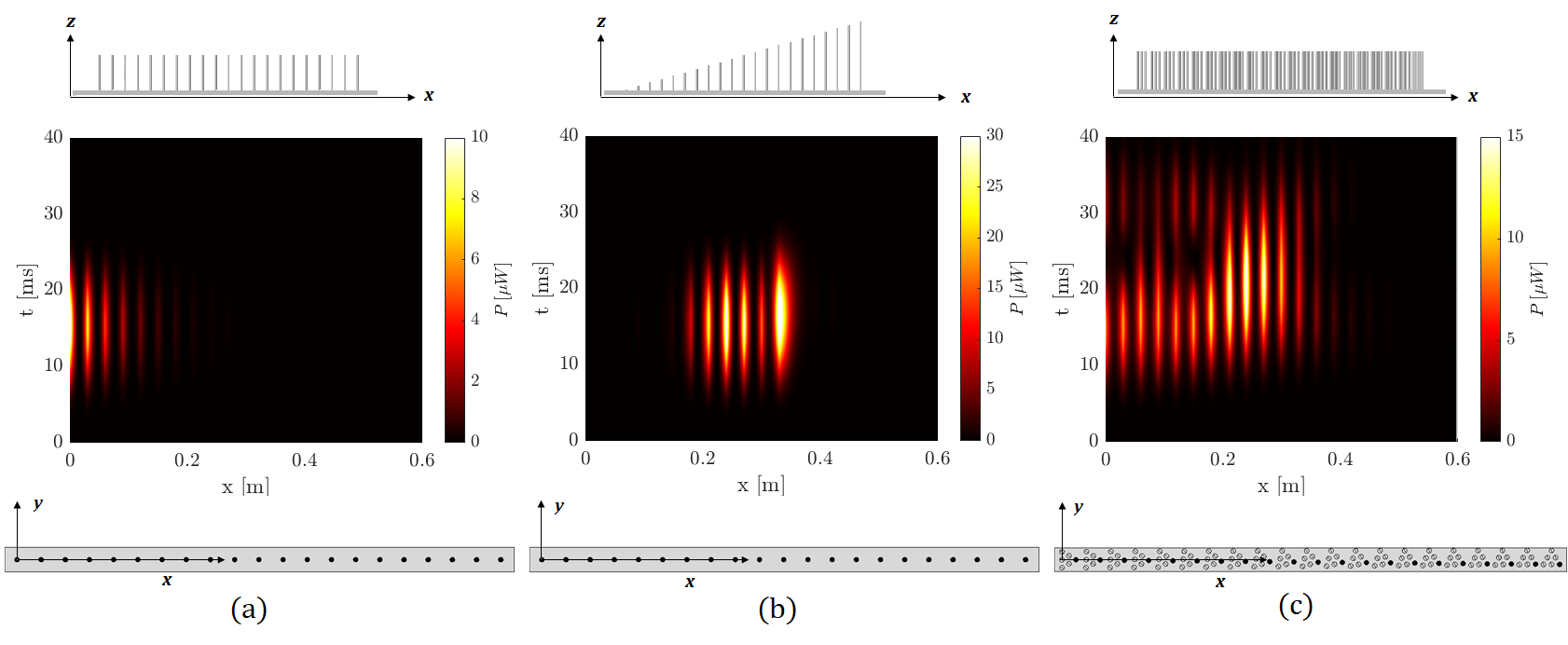}
    \caption{Spatiotemporal power for the ungraded array (a) the metawedge (b) and rotated cell array (c), corresponding to total scattering, rainbow reflection and rainbow trapping (input stops at $30\si{\milli\second}$). The arrays are excited through an $A_0$ Lamb mode at $13.5\si{\kilo\hertz}$. Each piezo disk is connected to a resistive load $R = 1 \si{\kilo}\Omega$.}
    \label{fig:Power}
\end{figure}

The reduction of reflection in the array with rotated cells is due to the bandgap opening in a position far from the edge of the Brillouin Zone, as outlined in Section~2; this provides a longer interaction of the wave with the resonators, since the coupling to a reversed wave is less than in the case where the grading introduces reflection from the band edge. In order to quantify the degree of trapping and the piezoelectric energy harvesting benefit, we consider a piezoelectric disk, of $3\si{\milli\meter}$ diameter and $2\si{\milli\meter}$ thickness, at the base of each resonator for each cases of the ungraded array and the rainbow reflective metawedge configuration. To ensure a fair comparison between the symmetry broken triangular configuration, only one piezoelectric disk is considered per cell (highlighted by red piezopatch in Fig.~\ref{fig:schem} and in aerial views in Figs.~\ref{fig:10msPropagation},\ref{fig:30msPropagation}). This ensures for the three cases, the amount of piezoelectric material is exactly the same. We then look at the spatiotemporal voltage output per base acceleration normalized with respect to the gravitational acceleration $g$. As shown by the dispersion bands in Fig. \ref{fig:10msPropagation} and \ref{fig:30msPropagation}, the unit cell for cases (a), (b) and (c) have been properly designed to obtain axial elongation of the rod with the piezo disk at the base at the frequency corresponding to the bandgaps opening ($13.5\si{\kilo\hertz}$). Thus, we have a dominant $d_{33}$ component of the  piezoelectric tensor. The piezoelectric material is PZT-5H with piezoelectric coefficients $d_{31}=-274\si{\pico\meter\volt^{-1}}$, $d_{33}=593\si{\pico\meter\volt^{-1}}$, $d_{15}=741\si{\pico\meter\volt^{-1}}$ and constant-stress dielectric constants $\epsilon_{11}^T/\epsilon_0 = 3130$ and $\epsilon_{33}^T/\epsilon_0 = 3400$ with $\epsilon_0 = 8.854\si{\pico\farad\meter^{-1}}$ the free space permittivity \cite{Erturk2011}. In order to estimate the electric power producible by the piezo-augmented arrays, each piezoelectric disk is attached to a resistive load $R = 1\si{\kilo}\Omega$. Piezo disks are electrically independent (no series or parallel connections) in order to avoid possible charge cancellation due to out of phase responses. This is numerically modeled using Abaqus complemented with a customised Fortran subroutine as in \cite{DePonti2019}. Computing the electrical power, we see the maximum local value is obtained by the rainbow \textit{reflection} array (Fig. \ref{fig:Power} (b)) with values approximately up to $30\mu\si{\watt}$. However, inspecting the time duration of the electric power production reveals that rainbow \textit{trapping} has the longest period of power output (Fig. \ref{fig:Power} (c)). The scattering array, i.e. the ungraded case has the worst performance as expected in both power produced and time duration (Fig.~\ref{fig:Power}(a)).

\begin{figure}
\hspace*{-0.2in}
    \centering
    \includegraphics[width = 1\textwidth]{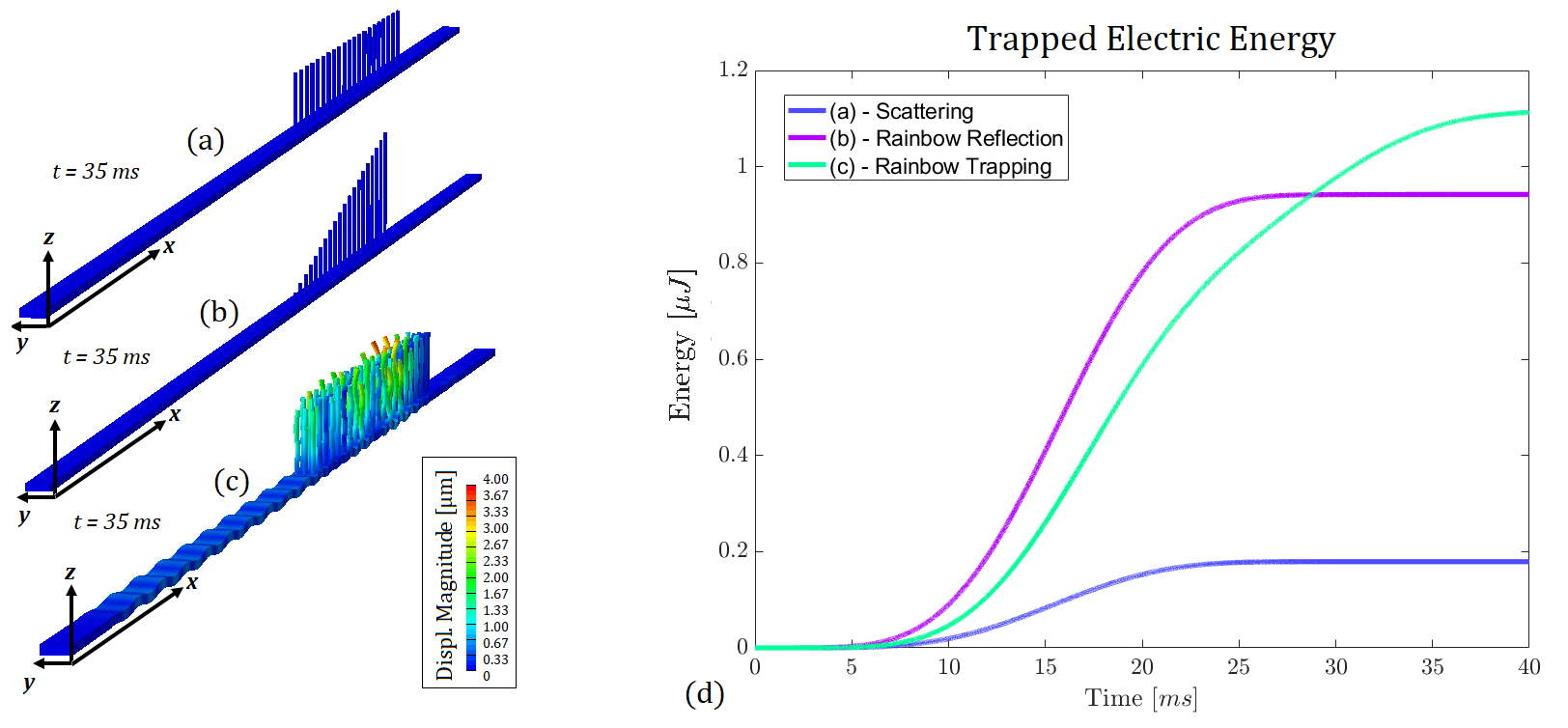}
    \caption{Displacement field along the arrays at time $35$ $ms$ for the ungraded (a) and the graded (b), (c) cases. Total electric energy stored in the arrays for increasing time (input stops at $30$ $ms$). See Supplementary videos for a visualisation of the reflection and trapping.}
    \label{fig:Energy}
\end{figure}

We then quantify the total energy trapped in the arrays, integrating along time the power produced by each piezo disk and summing all the obtained values. It can be seen that for the trapping case, the energy remains inside the array for a longer period of time (Fig. \ref{fig:Energy}), resulting in the highest trapped energy after approximately $28\si{\milli\second}$ (Fig. \ref{fig:Energy}). The total trapped energy at $40\si{\milli\second}$ is $0.18\mu\si{\joule}$, $0.94\mu\si{\joule}$ and $1.11\mu\si{\joule}$ for the ungraded, metawedge and rotated cells respectively. 

Therefore, by utilising rainbow trapping over rainbow reflection, it is possible to harvest more energy along the array, even though the largest local power was achieved by the reflective array; the simplicity of this structure allows the input mode shapes to match the modes of the single rods more effectively than in the symmetry broken arrays. The trapping arrays overcome this apparent downfall through the length of time the energy is trapped along the array, due to the lack of coupling to the reversed waves. Further to this, since the unit cells are more complex (i.e. more rods per cell) there is further scope for including larger amounts of piezoelectric material without compromising the resonances of the rods, e.g. at the base of every rod. 

\section{Conclusions and Perspectives}
We have delineated rainbow trapping and rainbow reflection effects for graded systems based on discrete unit cell structures, both with and without resonant elements. The distinction is not limited to elastic wave systems and  applies to all areas of wave physics where the slowing of wave propagation along a graded array is designed through locally periodic dispersion curves, be it in elasticity, electromagnetism, acoustics or water waves systems. As such, the term rainbow trapping should be limited for the original negative index materials \cite{Tsakmakidis2007}, or for describing systems which contain zero group velocity modes within the first BZ, often achieved by breaking the array symmetry, resulting in the lifting of accidental degeneracies in the dispersion relation. All other devices which manipulate the lowest dispersion curves achieve a rainbow reflection effect due to local Bragg scattering along the array; this can be achieved through the grading of any periodic structure although it is most efficient when resonant structures are used.

The implications of designing structures with these capabilities has been shown through the avenue of energy harvesting; ultimately true rainbow trapping devices gather more energy along the array due to the larger time period in which the array is localised at the trapping position. Coupled with this is the ability to include larger volumes of piezoelectric material due to the more complex geometries required. However, the simplicity of metawedge structures is not to be overlooked when designing energy harvesters. Due to their simplicity large local amplification of electrical power is achievable due to strong coupling with the incident radiation.

Due to the simplistic design paradigm for each effect, optimisation of the array parameters can be achieved enabling a broadband energy harvesting device; utilising both effects over a range of frequencies permits robust energy harvesting, over all areas of wave physics, which we envisage will lead to experimental verification.

\section{Acknowledgements}
The authors would like to thank Ian Hooper and Alastair Hibbins for useful conversations. GJC \& RVC acknowledge the UK EPSRC for their support through Programme Grant EP/L024926/1 and a Research Studentship. JMDP thanks Politecnico di Milano for the scholarship on "Smart Materials and Metamaterials for industry 4.0".

\section*{References}

\providecommand{\newblock}{}

\end{document}